# Two new parent compounds for FeSe-based superconducting phases

Shifeng Jin,[1,2] Xiaozhi Wu,[1] Qingzhen Huang,[3] Hui Wu,[3] Tianping Ying,[4] Xiao Fan,[1] Ruijin Sun,[1] Linlin Zhao,[1] and Xiaolong Chen[1,2,4]*

[1] *Research & Development Center for Functional Crystals, Beijing National Laboratory for Condensed Matter Physics, Institute of Physics, Chinese Academy of Sciences, Beijing 100190, China*

[2] *School of Physical Sciences, University of Chinese Academy of Sicences, Beijing 100190, China*

[3] *NIST Center for Neutron Research, National Institute of Standards and Technology, Gaithersburg, Maryland 20899, USA*

[4] *Laboratory of Advanced Materials, Fudan University, Shanghai 200438, China*

[5] *Collaborative Innovation Center of Quantum Matter, Beijing 100190, China*

It is well established that the occurrence of superconductivity in iron pnictides is closely related to the tetragonal to orthorhombic structural and antiferromagnetic (AFM) phase transitions. It, however, has not been clear whether the same scenario is appropriate for iron chalcogenide counterparts due to the absence of parent compounds for the latter family of superconductors. Here, we report the synthesis and structure determination of two novel phases in ethylenediamine intercalated FeSe, one is tetragonal and the other orthorhombic in room temperature, which can be stabilized with neutral spacer layers. Both phases can be regarded as the parent compounds for superconductivity as they are non-superconducting (non-SC) in pristine form and superconducting (SC) with $T_c$ up to 38K and 46K, respectively after Na doping, and the switch between SC and no-SC is reversible. Moreover, the two non-SC parent compounds show no evidence of long-range magnetic ordering down to 2K, only with dynamic spin fluctuations at low temperatures, suggesting that no competition between SC and AFM ordering. Our results reveal that undoped iron selenides are quantum paramagnetic in ground state, implying that they are distinct from their pnictide counterparts in pairing mechanism.

**Introduction:**

The recent discovery of various superconducting iron selenides with the highest $T_c$ above 100K [1,2,3,4] and novel electronic structures distinct from iron pnictides [5,6] has inspired a new fever in the investigation of unconventional superconductivity in FeSe-based compounds. As a key first step to unveil the nature of the superconducting phase, it is crucial to understand the normal state from which superconductivity arises. In iron pnictides, where numerous parent compounds exist, it is well-established experimentally that superconductivity is in the proximity of two ordered states that affecting the SC pairing, i.e. antiferromagnetic SDW order and nematic instability.[7,8] However, iron selenides is in extreme lack of parent compounds, which has roused intensive current debates on the nature of their parent states. As the only bulk selenide close to stoichiometry, the slightly electron doped $Fe_{1+\delta}Se$ ($\delta=0.01\sim0.04$) exhibits a 'nematic' transition below 90 K, however, the system shows no tendency towards magnetic ordering and becomes superconducting at 8.5K.[9] Based on the low-energy spin fluctuations observed in NMR measurements, it is previously suggested that $Fe_{1+\delta}Se$ should resemble electron-doped FeAs superconductors and be on the verge of an SDW ordering.[10] This scenario suggests an unified physics in pinctides and selenides superconductors and is further supported by the SDW like behavior of the multi-layer FeSe films observed in photoemission,[11] a static AFM order seems to be detected by μSR measurement in $Fe_{1+\delta}Se$ under high pressure,[12] and DFT calculations that suggest collinear SDW order in stoichiometric FeSe.[13] However, recently a growing number of theoretical calculations suggest the magnetic interactions in FeSe, as opposed to pnictides, demonstrate an unusual strong frustration, which may lead to new ground states including 'nematic quantum paramagnetic phase'[14] or 'antiferroquadrupolar ordering states'.[15] Furthermore, the absence of localized in-gap states in monolayer FeSe also seems incompatible with its proximity to magnetism. The increased contradictory results reported so far based on superconducting $Fe_{1+\delta}Se$ have made the parent states of iron selenides even more mysterious. It is therefore desirable to develop new strictly undoped FeSe-based compounds to reveal the parent state of this intriguing class of iron based superconductors.

Unlike the arsenides, the undoped iron selenide systems requires exclusively neutral constituent layers, which offer much less chemical stability because of the lack of interlayer Coulomb interaction. Instead, the appearance of interstitial cations or positively charged spacer layers are universal in the yet discovered iron selenides, which would inevitably induce superconductivity or, in some cases, ruin the stoichiometry of FeSe layers. In addition to $Fe_{1+\delta}Se$, several high-$T_c$ iron selenides were recently obtained by intercalating FeSe with K cations, molecules or LiOH as the spacer layer. In the case of $K_xFe_{2-y}Se_2$, the heavy electron doping expelled Fe ions at high temperature and resulted in minor SC phase that intergrowth with an AFM vacancy ordered phase.[16] The high-$T_c$ iron selenides intercalated with $NH_3$ in between stoichiometric FeSe layers[17,18] were only stabilized with co-intercalated dopants at some invariable concentration levels,[19] while several attempts to intercalate alkali metals with more heavier organic molecules yield ill-crystallized samples with unidentified structures and compositions.[20,21,22] Similarly, the recently reported $(Li_{0.8}Fe_{0.2})OHFeSe$ superconductor also request excessive metal cations (~20% $Fe^{2+}$) within the charged spacer layer,[23] leaving the iron vacancies in FeSe layers as the only significant compositional variable.[24] A challenge then arises for FeSe-based superconductors as how to strengthen the stability of the spacer layer, so that the parent compound is stable enough and eventually support a wide range of dopant concentration. Herein, we report on the synthesis of two novel phases in ethylenediamine (*En*) intercalated FeSe, *I* & *C* – $Na_x(C_2N_2H_8)Fe_2Se_2$, by means of sonochemistry & redox reaction, as well as direct solvothermal methods. Structure determination shows the *I*-phase adopts an *I*4/*m* tetragonal structure, while the *C*-phase directly synthesized by FeSe and *En* adopts a *Cmcm* orthorhombic structure. Strikingly, both phases can be stabilized with only neutral spacer layers and stay in non-SC parent states, while superconductivity can be readily introduced by Na intercalation, with $T_c$ up to 46K and 38K, respectively. Moreover, the magnetic susceptibility and heat capacity of the non-SC parent compounds show no evidence of long range magnetic order. The spin-disordered state appeared on cooling show highly degenerate states with an entropy plateau and exhibited gapless, linearly dispersive modes, supporting a frustration induced quantum paramagnetic ground state in the iron selenides.

**Results:**

The phase pure *I*- $Na_{0.5}(C_2N_2H_8)Fe_2Se_2$ intercalated by *En* molecules together with alkali metal was prepared by sonochemical solution reaction method (see Methods). Compared with the low-temperature solution-based route, the use of ultrasound in solution provides additional activation.[25] The synthesis yields a dark black product with composition $Na_{0.5}(C_2N_2H_8)Fe_2Se_2$. X-ray powder diffraction (PXRD) data showed no evidence for impurities within the instrumental resolution and the diffraction peaks were indexed on an *I*-centered tetragonal unit cell with lattice parameters $a$= 3.8145(7) Å and $c$ = 22.1954(8) Å at room temperature (Table 1). The crystal structure of *I*-$Na_{0.5}(C_2N_2H_8)Fe_2Se_2$ was resolved by *ab initio* structure determination from neutron powder diffraction (NPD) data (Figure 1). For solving this structure, ethylenediamine molecules and FeSe layers were treated as independent motifs in the simulated annealing approach, a preliminary structural model is built up with space group *I* 4/*m* and then Na position is located by Fourier difference analysis. The final Rietveld refinements produced an excellent fit to the NPD pattern, with $R_p$ = 5.87% and $R_{wp}$ = 7.66%.[26] The structure model determined from NPD is further confirmed by Rietveld refinements against X-ray diffraction data ($R_p$ = 5.72% and $R_{wp}$ = 7.32%). Rietveld fits and tables of the refined parameters are given in Fig. 1c, d and in the Supplementary Information (Table S1). As shown in Figure 1a, the structure resembles the reported crystal structure of $Li_{0.6(1)}(ND_2)_{0.2(1)}(ND_3)_{0.8(1)}Fe_2Se_2$ where the intercalated *En* molecules have strong orientational disorder. The molecules are located on general crystallographic positions 16i site (x; y; z), the resolved orientational disorder is easily understood by considering the 4$^{th}$-fold rotation axis and a mirror plane in between the *I*-centered FeSe layers. Na ions were located at 2a site (0, 0, 0) beside the *En* moieties. For clarity, Figure 1b shows the details of the structure by omitting the superimposed orientational disorder. The four N-H bonds of about 1 Å in the terminals of *En* molecules are all directing towards the selenide ions, with H- Se distances of 2.125 Å, and 2.480 Å respectively, in consistence with hydrogen bonding interactions found in the ammonia intercalates of FeSe and $TiS_2$. At ambient temperatures, the extremely compressed $FeSe_4$ tetrahedra found in $Fe_{1+\delta}Se$ are retained with Se-Fe-Se bond angles of 104.18(2)° and 112.18(3)° compared with values of 103.9(2)° and 112.3(4)° found in $Fe_{1+\delta}Se$. The Fe-Se bond distances are 2.3958 Å in $Fe_{1+\delta}Se$ and they are significantly larger in this intercalate, reaching 2.423(2) Å in $I$-$Na_xEnFe_2Se_2$ at ambient temperature, consistent with electron doping of the system. Meanwhile, the Se-Fe-Se bond angles of 106.64 (2)° and 110.91(3)° at 3 K indicate a decrease in the deformation of the $FeSe_4$ tetrahedra at lower temperatures and the Fe-Se bonds correspondingly contract to 2.343 (1) Å.

Figure 2a shows the temperature dependence of the magnetic susceptibility under an external magnetic field of 40 Oe. The as-synthesized $I$-$Na_{0.5}(C_2N_2H_8)Fe_2Se_2$ shows a diamagnetic transition at $T_c$ up to 46K with a considerable shielding fraction of 39% at 10 K in the zero-field-cooling process, indicating bulk superconductivity. Further confirmation of superconductivity is shown in the right inset of Figure 2a, which displays a typical magnetic hysteresis curve for a type-II superconductor, with its lower critical field $H_{c1}$ around 0.1 T. The divergence of the ZFC and field-cooled curves and the small ferromagnetic hysteresis above $T_c$ showed the presence of a ferromagnetic impurity corresponding to about 1~2% by mass of elemental Fe, which can be ascribed to the interstitial Fe in $Fe_{1+\delta}Se$ that reductively extruded in the reducing solution, as previously observed in $Li_{0.6(1)}(ND_2)_{0.2(1)}(ND_3)_{0.8(1)}Fe_2Se_2$.[18] Figure 2b shows the temperature-dependent electrical resistivity under a variety of magnetic fields, measured on a cold-pressed pellet. Apart from the metallic normal state above the $T_c$, a rapid decrease of resistivity was observed around 46 K without external field, and zero resistivity was reached at 40 K. The $T_c$ onset decreased continually with increasing magnetic field, the upper critical magnetic field can be determined by $H_{c2}(0) = 0.693[-(dH_{c2}/dT)]_{Tc}T_c$, which yields a high value up to 132 T.[27]

To extract the intercalated Na dopant from $I$-$Na_{0.5}(C_2N_2H_8)Fe_2Se_2$, the sample is exposed directly under air atmosphere and the redox reaction is investigated by an *in-situ* PXRD experiment with 10min intervals, as shown in Figure 3. No new peaks other than the $I4/m$ structure were observed during the air exposure, indicating no decomposition or phase transition occurs in the sample. Meanwhile, a closer view of 2θ = 8.5° ~ 10.5°, 16° ~ 38° portions indicates that peaks relating to (00l) shift obviously towards high-angles, others are nearly unchanged, reflecting a shrinkage of the unit cell in *c* axis. In particular, the shift of the (002) peak suggest the unit cell shrinkage occurs quickly at the first 60 mins and then relaxes with time. The observed phenomena can be attributed to a solid-solution redox reaction, evidenced by a continuous peak shift and lattice constant variation, with the structure only slightly altered without breaking $I4/m$ symmetry. All these structural changes indicate that $I$-$Na_{0.5}(C_2N_2H_8)Fe_2Se_2$ possesses a very stable structural framework. For comparison, the ammonia intercalated $Na_{0.65}(NH_3)_yFe_2Se_2$ phase easily decomposes at room temperature even under inert conditions ($O_2$ and $H_2O$ concentration,<0.1 p.p.m.), which is attributed to the deintercalation of $NH_3$ molecules from the interlayers.[28]

Figure 4 shows the PXRD pattern for an oxidized product of $I$-$Na_{0.5}(C_2N_2H_8)Fe_2Se_2$ sample that exposed in air over two weeks. All reflections were indexed using an *I*-centered tetragonal unit cell with lattice parameters of $a$ = 3.8565(2) Å and $c$ = 21.4257(6) Å. Systematic absences were consistent with the space groups *I*4/*m*, and the structure determination resulted in a structural model shown in Figure 4. This new model, with only disordered **En** molecules and free of dopant Na within the FeSe interlayers, giving an excellent Rietveld fit to the PXRD pattern (Figure 4, Tables S2). The refined structure of *I*- $(C_2N_2H_8)Fe_2Se_2$ manifests the deintercalation of Na from *I*-$Na_{0.5}(C_2N_2H_8)Fe_2Se_2$ and slightly titled positions of **En** molecules, confirming the structure affinity of the two compounds. Meanwhile, the decreased Fe−Se distances of 2.3741(2) Å and the much less deformed $FeSe_4$ tetrahedra with Se−Fe−Se angles of 109.893(1)° (×2) and 108.631(1)° (×4) in comparison with *I*-$Na_{0.5}(C_2N_2H_8)Fe_2Se_2$ are consistent with the removal of electron doping. Interestingly, the results indicate that only Na atoms were significantly de-intercalated from the interlayers during air exposure, and the compound persists its structure even without dopant. *I*-$(C_2N_2H_8)Fe_2Se_2$ can then be regarded as a parent compound of FeSe-based superconductors that with charge neutral space layers. Magnetometry on the same batch of materials used for the in situ PXRD measurements revealed decreasing in both $T_c$ and superconducting shielding fraction along with Na deintercalation, with no evidence of superconductivity after 60 mins. The susceptibility results support that the metal content (doping level) plays a decisive role in determining the value of the $T_c$, in agreement with the recent theoretical calculations by Guterding et al.[29] Figure S1 shows the specific heat measured between 2-175K for *I*-$(C_2N_2H_8)Fe_2Se_2$, which indicates the absence of long range magnetic ordering down to 2K. In addition, Na atoms can be restored to the deintercalated compound *I*-$(C_2N_2H_8)Fe_2Se_2$ through additional sonochemical solution reaction with Na pieces, and superconductivity with $T_c$ up to 45 K can be recovered in a compound with lattice parameters comparable to *I*-$Na_{0.5}(C_2N_2H_8)Fe_2Se_2$ (Figure S2).

The superconductor *I*-$Na_{0.5}(C_2N_2H_8)Fe_2Se_2$ and its parent compound *I*-$(C_2N_2H_8)Fe_2Se_2$ are found thermally unstable, i.e., heating the samples under vacuum or in air below 100 °C was sufficient to initiate an irreversible phase transition. Temperature dependence of the diffraction profiles for *I*-$(C_2N_2H_8)Fe_2Se_2$ was subsequently measured from 27°C to 200°C. As shown in Figure 5, the *in situ* PXRD measurements of the *I*- $(C_2N_2H_8)Fe_2Se_2$ initially showed discrete reflections due to high symmetry of the compound. At a higher temperature (80°C), some reflections in the pattern rapidly diminished and were replaced by a more complex set of reflections. These new reflections bear a limited resemblance to those of the previous *I*4/*m* $(C_2N_2H_8)Fe_2Se_2$ in which the two lowest-angle Bragg peaks corresponding to (002) and (004) reflections are almost unaltered, with a d-spacing of the (002) reflection ∼10.8 Å. However, the appeared (hkl) reflection conditions probed *in situ* substantially changed from $h+k+l=2n$ to $k+l=2n$, corresponding to a fundamental change in extinction conditions from *I*-centered to *A*-centered. Further increasing temperature above 80°C, the Bragg peaks due to *I*4/*m* $(C_2N_2H_8)Fe_2Se_2$ were completely replaced by those of the *A*-centered new structure, which is an irreversible phase transformation and do not undergo any further changes upon cooling. The diffraction profiles of the new product persist up to 200 °C, with a slight decomposition occurring higher than 180°C (Figure 5a). Interestingly, so far a majority of the reported iron based superconductors are *P* or *I*-centered, while *A*-centered iron based superconductors have not been reported.[30] Figure 5b schematically represents the square anti-fluorine type FeAs/FeSe layers, where the spatial arrangement of adjacent FeAs/FeSe layers have three symmetric configurations, i.e., *P*, *I* and *A* types. For the more frequently encountered *P* and *I* centered iron based superconductors, the interlayer environment for space layer is isotropic along *a* and *b* directions. However, as shown in figure 5b, a (0, 1/2, 1/2) displacement of the FeAs/FeSe corrugated layers leads to a different arrangement of these layers along the two basal directions, the interlayer environment for an *A*-centered iron based superconductor is then anisotropic.

To fully describe the structure of the new *A*-centered intercalate, a new solvothermal synthesis route is adopted to produce high crystalline materials for accurate structure characterizations (see methods). High purity *A*-centered material can be directly synthesized from *β*-FeSe and pure **En** in the temperature range between 150 ~ 220°C, without the presence of any alkali metal. The elevated reaction temperature and pressure in the autoclaves up to 220 °C or addition of appropriate mineralizer (e.g. KCl, $NH_3Cl$) can improve the crystallinity and lead to shining crystalline products with metallic luster. The products in composition $(C_2N_2H_8)Fe_2Se_2$ were stable in air and appeared single phase examined using high resolution NPD data (Figure 6). All reflections at room temperature were indexed using an *orthorhombic* cell with lattice parameters of *a* = 3.8607(2) Å, *b* = 3.8926(2) Å and *c* = 21.6618(4) Å, with systematic absences in consistence with the space groups *Ama*2, *Amma* (*Cmcm*) or *A*2122, etc. A subsequent *ab-initio* structure solution in space group *Cmcm* successfully resulted in a structural model with alternated FeSe and disordered **En** molecular layers that extend along *c* axis by a *c* glide plane. The disordered **En** molecules zigzagged along *c* direction and well matched the anisotropic interlayer vacant. This new model, with charge neutral space layers, gave an excellent Rietveld fit to the high resolution NPD pattern, with $R_p$ = 5.46% and $R_{wp}$ = 7.64%. The structure with refined composition $(C_2N_2H_8)Fe_2Se_2$ is shown in Figure 6 a, b & c, and the refined parameters are supplied in Tables S3. The Fe and Se atoms were located on crystallographic positions 4*c* (0,*y*,1/4) and 4*b* (0, 1/2,0) sites, while **En** molecules resided on the 16*h* (*x*,y,z) general sites, respectively. The orientational disordered **En** molecules were aligned so that their C-C bonds almost parallel to the *b* axis, while the C-N-C planes are almost parallel to the *bc* plane. The anisotropic orientation of **En** molecules explained the slight deformation of the cell between *a* and *c* axes. The deformation of FeSe layers in ***C***-$(C_2N_2H_8)Fe_2Se_2$ is small compared with those previously reported FeSe based intercalates, with Se-Fe-Se bond angles 110.61(1)° (×4), 107.43(1)° and 106.98(1)° at 3K, respectively. Meanwhile, the Fe-Se bond distances of 2.405(3) Å (×2) and 2.392(2) Å (×2) in ***C***-$(C_2N_2H_8)Fe_2Se_2$ are longer than those in the *I*4/*m* **En** intercalates.

The synthesis route of ***C***-($C_2N_2H_8$)$Fe_2Se_2$ precludes the possibility of contamination from alkali metal dopant, which supplied a rare chance to probe the clean parent state of FeSe based superconductors. The magnetic susceptibility ($\chi$) measurements under an external field of 40 Oe and 1 T for ***C***-($C_2N_2H_8$)$Fe_2Se_2$ are shown in Figure S3 and Figure 7a, respectively. No superconducting signal was observed down to 2K, instead, the temperature-dependent magnetic susceptibility exhibits a Curie-Weiss behavior above 40 K: $\chi=C/(T-\theta)$. The Curie constant C corresponds to an effective moment of 1.92(1) Bohr magnetons ($u_B$), and the negative Weiss temperatures -5.7 K indicates week AFM interactions. No sharp magnetic anomaly or field hysteresis was observed down to 1.8 K. Instead, the susceptibility smoothly increased on cooling and passed through a shallow and broad maximum at T =10 K before approaching a finite low-T limit of 0.3 $u_B$/Fe in external field of 9T. The saturate moment is comparable to the SDW moment revealed for FeAs based superconductors.

Specific heat data, $C_P(T)$, also showed no evidence of magnetic phase transition for fields between 0 T and 9 T and temperatures down to 0.3 K (Fig. 7B). These data indicate a disordered low-temperature state without conventional AFM order down to 0.3K. Interestingly, $C_P/T$ of ***C***-($C_2N_2H_8$)$Fe_2Se_2$ shows a broad entropy peak below 10 K, the $C_p/T$ vs. $T^2$ data (inset of figure 7b) also show significantly deviation from linear even down to 0.3 K. The entropy peak is field sensitive and can be significantly suppressed at 9T, indicating strong magnetic contributions to the $C_p$ besides phonon below 10 K. Meanwhile, corresponding to the $\chi$ plateau observed in 9 T, below 3 K a perfect $T^3$ form of $C_p$ is restored at 9 T, where under extremely high field the spins are frozen with lowering temperatures and the lattice contribution is well separated.

We then probed the density of states for spin excitations through the magnetic specific heat $C_M(T)$ after subtraction of the lattice contribution. The specific heat exhibited power-law behavior at low temperatures (Fig. 7B). The data between 0.3 and 3.0 K are well fitted by a power law $C_M = AT^\alpha$, where A is a constant and $\alpha$ is 2.110(5). Quadratic temperature dependence through one decade indicates the presence of gapless and linearly dispersive modes in two dimensions. A gapless spectrum is furthermore consistent with a finite value of the susceptibility in the low-T limit. Meanwhile, the $C_p/T$ data at 9 T also exhibited a broad peak structure in comparison with the zero field $C_p/T$ above 10K (Fig. 7B), which arises from the release of spins frozen by external field with elevating temperature and thus contributes to the increase of $C_m$. $C_p/T$ at 9T finally converge to the $C_p/T$ at 0T around 40K, where the system falls into the high-temperature Curie-Weiss law (Fig. 7A). Similar thermodynamic properties have been reported for magnetic frustrated low dimensional system $NiGa_2S_4$, which was proposed for the potential spin liquid systems due to its geometrical magnetic frustration. The ***C***- ($C_2N_2H_8$)$Fe_2Se_2$ differs from the $NiGa_2S_4$ in that the spin excitation in terms of field sensitivity, the linearly dispersive excitations implied by $C_M \propto T^2$ are then collective modes of spin clusters with net magnetic moments. The spin-disordered state of ***C***- ($C_2N_2H_8$)$Fe_2Se_2$ present on cooling shows highly degenerate states with an entropy plateau and exhibits gapless, linearly dispersive modes, supporting a frustration induced quantum paramagnetic ground state in the iron selenides.

Analysis of the solvothermally synthesized *C*- $(C_2N_2H_8)Fe_2Se_2$ samples shows that superconductivity is absent when the spacer layer is formally neutral. A subsequent reductive sodiation using sodium/ethylenediamine solution was applied to *C*- $(C_2N_2H_8)Fe_2Se_2$ for Na intercalation (see methods). As shown in Figure S4, the crystal structure of the *Cmcm* compounds was maintained at nominal compositions: $Na_x(C_2N_2H_8)Fe_2Se_2$ (x=0.0, 0.5, 1.0), only at x=1.0 additional reflections appeared from trace impurities. Magnetic properties of the samples were then investigated down to 10 K. Results shown in Figure 7d are the temperature dependence of their magnetic susceptibilities measured under 40 Oe after zero-field cooling. With the doping increased, the onset $T_c$ of *C*-$Na_x(C_2N_2H_8)Fe_2Se_2$ increased significantly from 28 K (x=0.5) to 38 K (x=1.0). In particular, the estimated superconductive shielding fraction of the last two samples reached about 30% at 10K, confirming bulk superconductivity appeared in the *Cmcm* phase. The fabrication of higher-quality *I*4/*m* phase and *Cmcm* phase crystals with a wider range of doping level and higher shielding fraction is underway so as to fully understand the superconducting phase diagrams.

**Discussion**

Hendricks pointed out that the organic bases are held by van der Waals forces in addition to Coulomb forces within the interlayer space.[31] Hence the larger *En* bases are more strongly absorbed, and successfully stabilized the neutral host lattice without coulomb forces between the charged host layers and the metal ion dopants. Beside the air stable *C*-$(C_2N_2H_8)Fe_2Se_2$, the enhanced stability of *En* intercalated compounds has also been demonstrated in air oxidation of *I*-$Na_{0.5}(C_2N_2H_8)Fe_2Se_2$, where only the Na ions are deintercalated while the neutral host lattice remains intact in air for months. The stable parent compounds reported here have unique superiority for in-depth physical property investigation and further superconductivity optimization .

The paramagnetic magnetic behavior of the undoped *I*-$(C_2N_2H_8)$ $Fe_2Se_2$ and *C*-$(C_2N_2H_8)Fe_2Se_2$ revealed in this study differs apparently from the magnetic ordered states in the undoped FePh superconductors. Previous investigations suggest that the parent phase for FeAs superconductors is magnetically ordered in a SDW state, and the SC emerges when carrier doping suppresses the SDW instability.[32] However, $Fe_{1.01}Se$ and its intercalates with metal dopants show no evidence of long range magnetic ordering and exhibit superconductivity between 8K~46K when the FeSe layers are close to their stoichiometry. Recently, Dong et al. reported for the first time an antiferromagnetic SDW analogue occurs at ~127 K in the non-SC FeSe-based samples of $(Li_{0.8}Fe_{0.2})OHFeSe$.[23] However, a detailed structure analysis in $(Li_{0.8}Fe_{0.2})OHFeSe$ system using neutron powder diffraction and X-ray absorption spectroscopy pointed out these non-SC samples have a wide range of Fe vacancy concentrations within the FeSe layers, and the spacer layer is still heavily charged by $Fe^{2+}$.[24] Thus, we infer from our experimental that the *strictly undoped* parent phases of FeSe-based superconductors are *paramagnetic*, other than the antiferromagnetism state proposed by early theoretical calculations.[33,34] Our finding is more consistent with more recent electronic structure calculations,[14] which concluded that a non-magnetic ground state is most stable for iron selenide due to magnetic frustration.[35] Moreover, $^{77}$Se NMR investigation of $Fe_{1.01}$Se found strong enhancement of antiferromagnetic spin fluctuations towards $T_c$, suggesting that $Fe_{1.01}$Se resemble electron-doped FeAs superconductors and could be on the verge of an SDW ordering. Further neutron scattering and NMR investigations on the *I*-$(C_2N_2H_8)Fe_2Se_2$ and *C*-$(C_2N_2H_8)Fe_2Se_2$ could be helpful in clarifying the possible antiferromagnetic fluctuations below 10 K in those *En* intercalated parent compounds.

On the other hand, the question of how the host structure parameters and doping influence $T_c$ in the intercalates of FeSe has been raised for years and yet not well answered. It is widely accepted that the ideal $FeAs_4$ tetrahedron is favorable for superconductivity in FeAs-based superconductors.[27] Recently, based on the heavily distorted $FeSe_4$ tetrahedron found in both structures of high-$T_c$ superconductors $Li_{0.6(1)}(ND_2)_{0.2(1)}(ND_3)_{0.8(1)}Fe_2Se_2$ (43K) and $(Li_{0.8}Fe_{0.2})OHFeSe$ (44K), Lu et al. proposed that $FeSe_4$ distortion is likely to play a key structural role in enhancing the superconductivity.[23] Nonetheless, the resolved crystal structures of the two parent phases of *En* intercalates all adopt near ideal $FeSe_4$ tetrahedrons, and after electron doping the superconducting phase features extended Fe-Se bonds and extremely compressed $FeSe_4$ tetrahedra. Then it seems $FeSe_4$ distortion is rather a consequence of electron doping, which naturally influenced the $T_c$. Meanwhile, the dependence of interlayer separation to the $T_c$ enhancement was discussed in pioneer studies on molecular intercalation, without concerning the exact doping levels. The interests on expanding FeSe interlayer have centered on the improved two-dimensionality of the band structure, which resembles the case in the monolayer FeSe that with the highest $T_c$. Recently Guterding et al. highlight in theory that $T_c$ should be mainly controlled by the electron doping levels,[29] not the sufficiently large FeSe interlayer distances. Further experimental work should therefore concentrate on fine controlling the doping level toward the optimum doping zone, a task previously impossible in FeSe based systems due to structure instability is now more promising based on the stable *En* based parent compounds and their future derivatives.

In conclusion, we successfully isolated two pure superconducting phases in ethylenediamine (*En*) intercalated FeSe system, *I* and *C*-$Na_x(C_2N_2H_8)Fe_2Se_2$, by means of novel sonochemistry and solvothermal methods. Structure solutions and refinements against NPD and PXRD data show that the ultrasoniced *I*-phase adopts a high symmetric *I* centered tetragonal structure, while the *C*-phase by solvothermal adopts a rare *C* centered orthorhombic structure. Both phases can be stabilized without metal dopant and represent the rare FeSe-based examples with charge neutral space layers. We demonstrate that high-quality samples of *I* and *C* -($C_2N_2H_8$)$Fe_2Se_2$, two bulk non-SC parent phases with only charge neutral intercalates, exhibit a spin disordered state in two dimensions. Despite antiferromagnetic (AFM) interactions in *C* -($C_2N_2H_8$)$Fe_2Se_2$, no magnetic long-range order was observed down to 0.3 K. Instead, the spin-disordered state appeared on cooling showed highly degenerate states with an entropy plateau and exhibited gapless, linearly dispersive modes, supporting a frustration induced quantum paramagnetic ground state in the iron selenides. Further, Na metal can be freely intercalated to these non-SC parent compounds and induced superconductivity with Tc up to 46K and 38K, respectively. The soft chemical methods described here has led to high crystalline FeSe-based superconductors and the previously non-existent parent compounds with various symmetries, and is sufficiently flexible for isolating new families of FeCh superconductors required for the exploration of the electronic states at previously inaccessible states.

**Methods:**

**Structural characterization and composition determination.** NPD experiments were conducted using Ge 311 (λ=2.0775 Å ) and Cu 311 (λ=1.5401 Å ) monochromators, respectively. Data were collected over the 2θ ranges 1.3-166.3° (Ge 311) and 3-168° (Cu 311) with a step size of 0.05° from 3K to 295 K. The NPD experiments were carried out in the NIST Center for Neutron Research. Laboratory *in situ* PXRD measurements were made using a Panalytical X'pert PRO instrument (Co Kα1 radiation) equipped with an Anton Paar HTK-1200N Oven Sample stage ($10^{-4}$pa, RT-1200°C). Time-dependent *in situ* X-ray diffractometry was performed in air atmosphere at intervals of 10 min (5° to 60°, with a scanning step width of 0.017°). Temperature-dependent *in situ* X-ray diffractometry was performed in the angular range from 5° to 65°. The room-temperature diffraction pattern was firstly obtained as a standard, the sample stage was then heated from 60°C to 200°C at intervals of 10°C. Each diffraction pattern was obtained 5 min after the required temperature was reached. The sample composition was determined by inductively coupled plasma mass spectrometry and thermogravimetry.

**Magnetic susceptibility, resistivity and heat capacity.** Magnetization, resistivity and heat capacity measurements were carried out using a SQUID PPMS-9 system (Quantum Design). Magnetic susceptibility measurements were made in *d.c.* fields of 40 Oe or 1 T in the temperature range 2-300 K after cooling in zero applied field (ZFC) and in the measuring field (FC). The low temperature heat capacity was measured on the cold-pressed powder sample by thermal relaxation method using PPMS in the temperature range 0.3-300 K. Resistivities were measured using the standard four-probe configuration based on samples cold-pressed at a uniaxial stress of 400 kg cm$^2$.

.

**Table 1.** Crystallographic parameters from the Rietveld refinement of *En* intercalates at 298 K.

| Composition | Symmetry | $a$ (Å) | $b$ (Å) | $c$ (Å) |
|---|---|---|---|---|
| Na$_{0.5}$(C$_2$N$_2$H$_8$) Fe$_2$Se$_2$ | *I* 4/*m* | 3.824(3) | - | 22.179(7) |
| (C$_2$N$_2$H$_8$)Fe$_2$Se$_2$ | *I* 4/*m* | 3.8565(2) | - | 21.4257(6) |
| (C$_2$N$_2$H$_8$)Fe$_2$Se$_2$ | *Cmcm* | 3.8779(2) | 21.359(2) | 3.8454(4) |

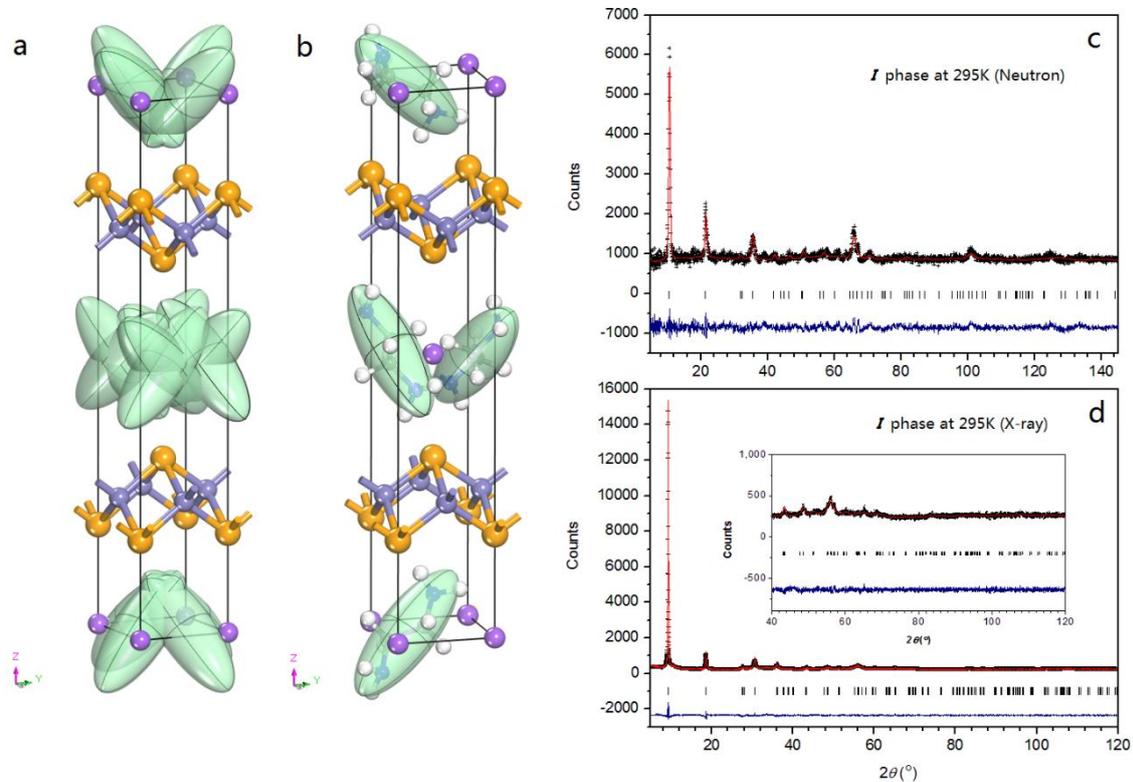

**Figure 1. Crystal structure and Rietveld refinement against Neutron & PXRD data for *I*- Na$_{0.5}$(C$_2$N$_2$H$_8$)Fe$_2$Se$_2$ at 295 K**. **a**, structure model with C$_2$N$_2$H$_8$ molecule represented as ellipsoids, orientational disorder of C$_2$N$_2$H$_8$ molecules are due to high symmetric local environment in between FeSe layers. **b**. structure details of **α** phase omitting the superimposed orientational disorder **c.** Refinement against Neutron (1.5396 Å) and X-ray(Co Kα) powder diffraction data **(d)**, observed data (crosses), calculated data (red solid line), differences (navy solid line), vertical bars ( | ) indicate the positions of the Bragg peaks, inset in **d** shows the high angle data in×5 scale.

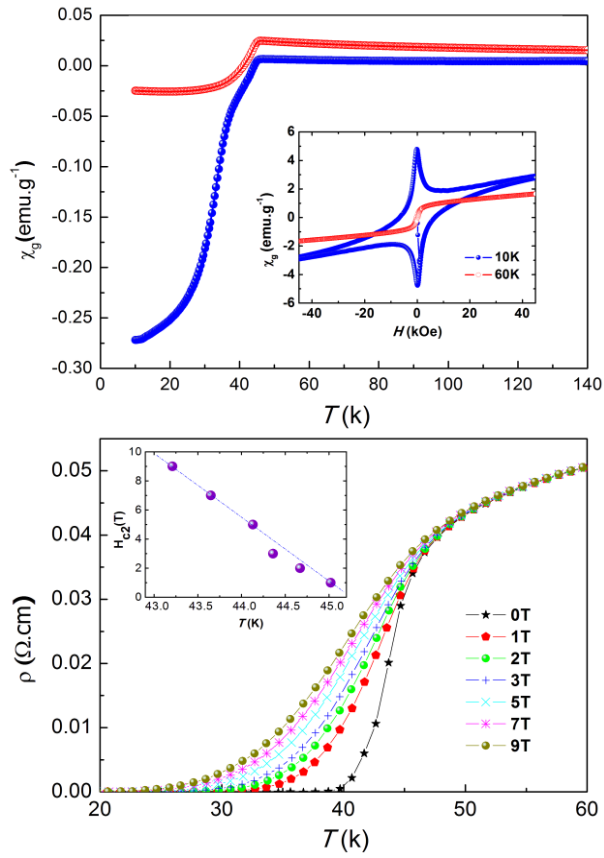

**Figure 2. Magnetization, electrical resistance of *I*-Na$_{0.5}$(C$_2$N$_2$H$_8$)Fe$_2$Se$_2$. a,** ZFC and field-cooled measurements on the sample of Na$_{0.5}$(C$_2$N$_2$H$_8$)Fe$_2$Se$_2$, inset shows the magnetic hysteresises of nominal NaFe$_2$Se$_2$ measured at 10 K and 60 K in the range -50 kOe to 50 kOe, respectively. **b**, Temperature dependence of electrical resistivity under magnetic fields of 0, 1, 2, 3, 5, 7, and 9 T, inset shows the temperature dependence of the upper critical magnetic field.

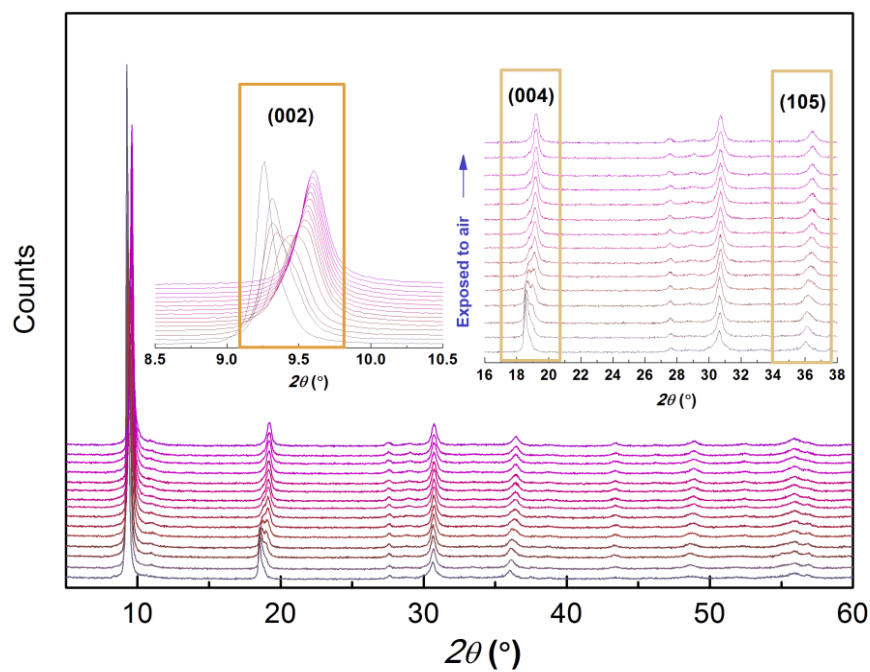

**Figure 3. Structure evolution of α-Na$_{0.5}$(C$_2$N$_2$H$_8$)Fe$_2$Se$_2$ on air oxidation.** In situ XRD patterns collected during the oxidation of Na$_{0.5}$(C$_2$N$_2$H$_8$)Fe$_2$Se$_2$ under air. Insets show continual shift of diffraction peaks relating to the shrinkage of the *c* lattice.

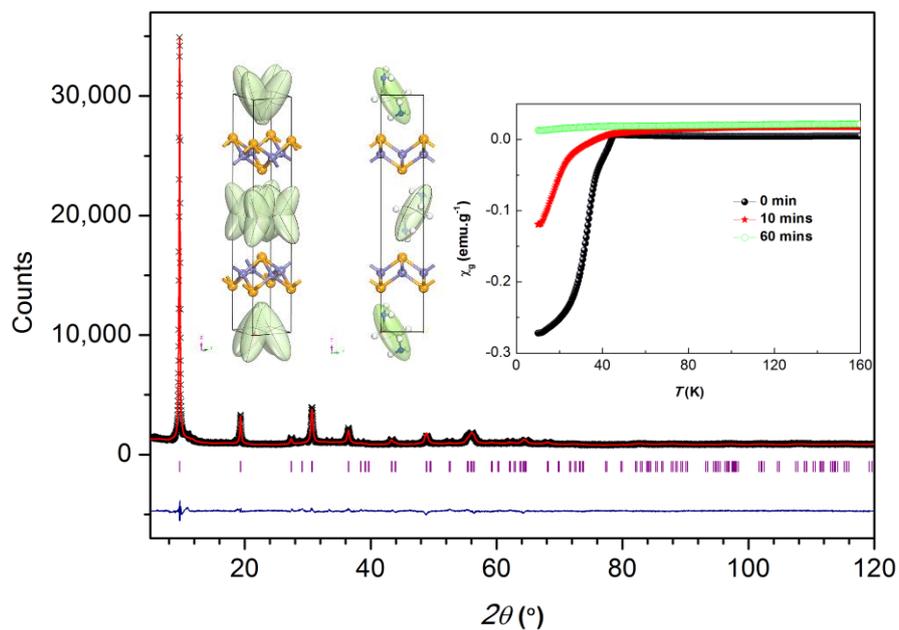

**Figure 4.** Observed (black crosses) and calculated (red line) PXRD pattern and difference profile of the Rietveld refinement of the structure of the oxidized sodium free **α-** (C$_2$N$_2$H$_8$)Fe$_2$Se$_2$ at 298 K and the final crystal structure (left inset), peak positions are marked by vertical lines. Right inset shows ZFC measurements on the oxidized sample of Na$_{0.5}$(C$_2$N$_2$H$_8$)Fe$_2$Se$_2$.

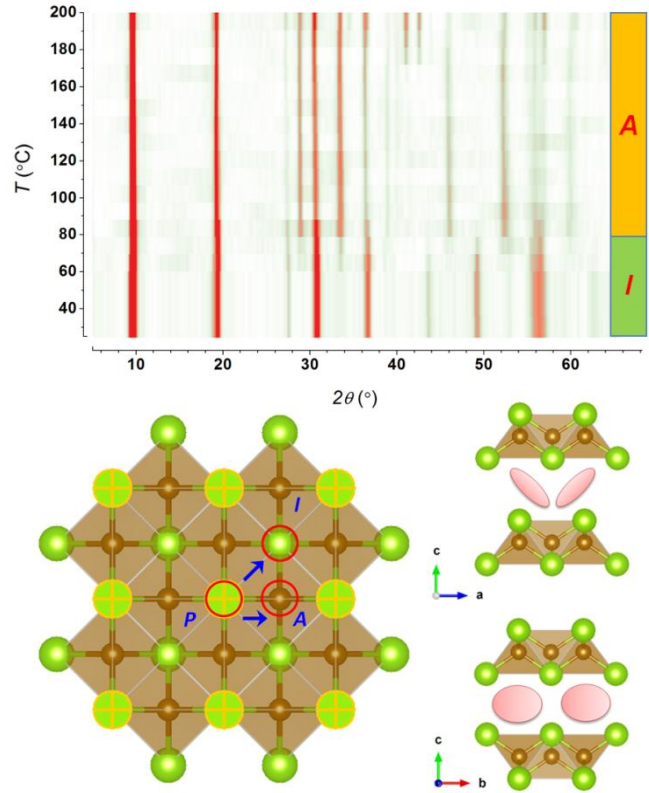

**Figure 5. Structure evolution of (C₂N₂H₈)Fe₂Se₂ with elevated temperature. a**, Film plot showing the *in-situ* high temperature XRD patterns; an invariable phase transformation occurred at 80 °C and corresponding to a new phase with *A* center. **b**, Schematic representation of relative positions of Se atoms in the adjacent FeSe layers for *P*, *I* and *A* centered compounds. Left insets shows the anisotropic interlayer vacant in between FeSe layers in an *A* centered compounds view along *a* and *b* directions.

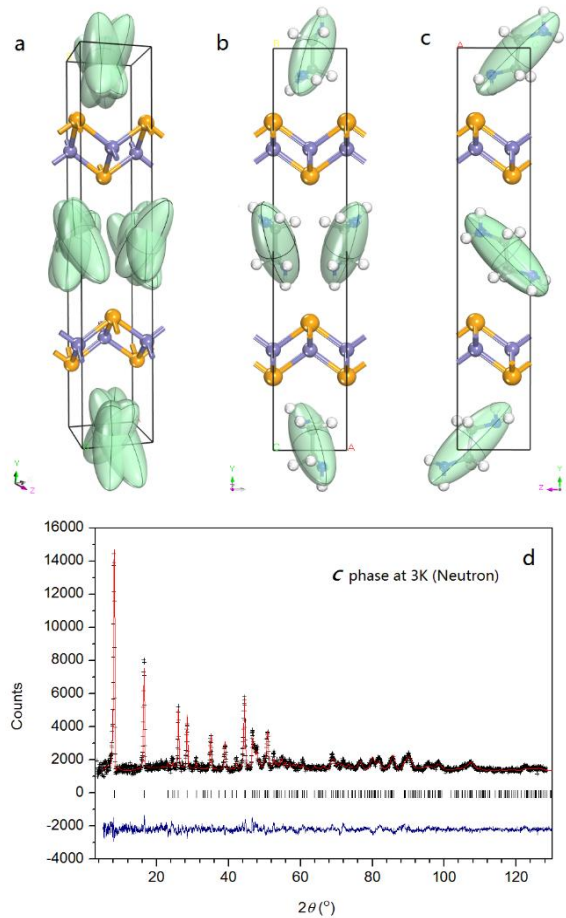

**Figure 6. Crystal structure and Rietveld refinement against NPD data for *C* centered *C*-(C$_2$N$_2$H$_8$)Fe$_2$Se$_2$. a**, The crystal structure model with C$_2$N$_2$H$_8$ molecule represented as ellipsoids, orientational disorder of C$_2$N$_2$H$_8$ molecules are superimposed within a unit cell. **b**. structure projected along the [001] direction and [100] direction **(c)**, the orientational disorder is omitted for clarity **d.** Refinement against Neutron (1.5396 Å) powder diffraction data, inset shows the high angle data in×5 scale, observed data (crosses), calculated data (red solid line) ,differences (navy solid line), vertical bars ( | ) indicate the positions of the Bragg peaks.

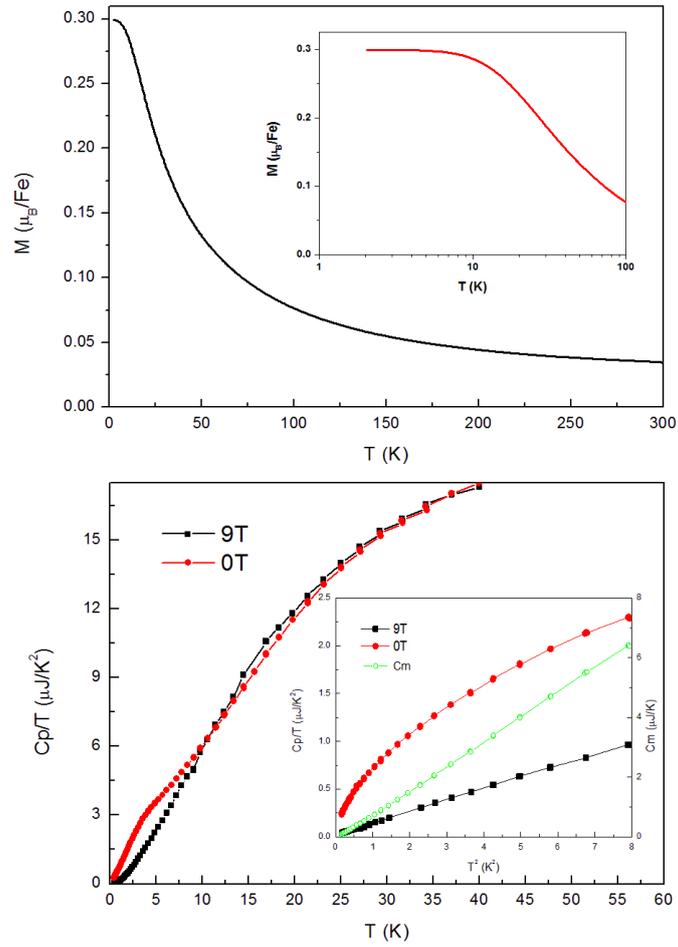

**Figure 7. Magnetic susceptibility and heat capacity measurements. a**, ZFC magnetic susceptibilities of ***C***-($C_2N_2H_8$) $Fe_2Se_2$ samples and Curie-Weiss fit. left inset, linear Curie-Weiss fit below 200K, **b**, heat capacity of cold-pressed ***C***-($C_2N_2H_8$)$Fe_2Se_2$ powder.

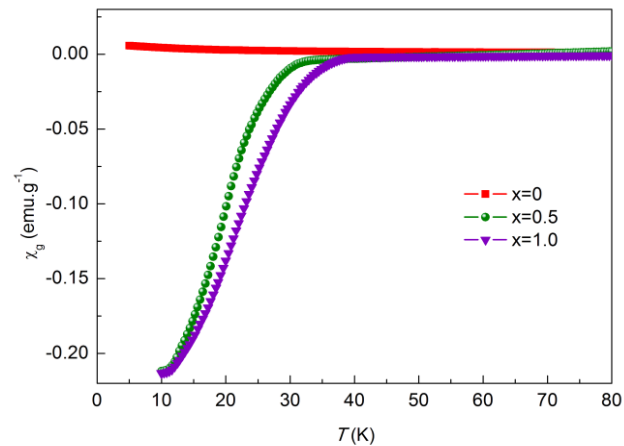

Figure 8. ZFC measurements on the nominal ***C***-Na$_x$($C_2N_2H_8$)$Fe_2Se_2$ ($x$=0,0.5,1.0).

# Supplementary Information

**Table S1. Structural Parameters**. Atomic coordinates for *I* - Na$_{0.5}$(C$_2$N$_2$H$_8$)Fe$_2$Se$_2$ from NPD refinement at 295K. R$_p$ = 5.87% and R$_{wp}$ = 7.66%.

| Atom | Wyckoff site | x | y | z | Occupation | $U_{iso}$(Å$^2$) |
|---|---|---|---|---|---|---|
| N1 | 16i | -1.173(8) | 0.959(8) | 0.418(9) | 1.0 | 0.114(3) |
| C1 | 16i | -1.332(6) | 0.983(7) | 0.478(8) | 1.0 | 0.114(3) |
| C2 | 16i | -1.732(6) | 0.992(8) | 0.480(9) | 1.0 | 0.114(3) |
| N2 | 16i | -1.859(5) | 1.021(9) | 0.543(9) | 1.0 | 0.114(3) |
| H1 | 16i | -1.167(9) | 1.202(9) | 0.397(14) | 1.0 | 0.114(3) |
| H2 | 16i | -1.316(15) | 0.796(16) | 0.390(14) | 1.0 | 0.114(3) |
| H3 | 16i | -1.234(7) | 1.223(8) | 0.502(9) | 1.0 | 0.114(3) |
| H4 | 16i | -1.237(6) | 0.755(8) | 0.506(9) | 1.0 | 0.114(3) |
| H5 | 16i | -1.829(7) | 0.756(9) | 0.455(9) | 1.0 | 0.114(3) |
| H6 | 16i | -1.826(8) | 1.228(9) | 0.455(9) | 1.0 | 0.114(3) |
| H7 | 16i | -1.754(12) | 1.242(9) | 0.563(10) | 1.0 | 0.114(3) |
| H8 | 16i | -1.790(13) | 0.805(9) | 0.569(12) | 1.0 | 0.114(3) |
| Se | 4e | 0.50000 | 0.50000 | 0.316(1) | 1.0 | 0.165(5) |
| Fe | 4d | 0.00000 | 0.50000 | 0.25000 | 1.0 | 0.171(4) |
| Na | 2a | 0.50000 | 0.50000 | 0.50000 | 0.35(7) | 0.08(5) |

Space group: *I*4/*m* (No. 86); *a* = 3.824 (3) Å, *c* = 22.179 (7) Å, *V* = 324.3(6) Å$^3$, $U_{iso}$ is the isotropic displacement parameter.

**Table S2. Structural Parameters**. Atomic coordinates for *I* - (C$_2$N$_2$H$_8$)Fe$_2$Se$_2$. *R$_p$* = 3.49% and *R$_{wp}$* = 4.57%.

| Atom | Wyckoff site | x | y | z | Occupation | $U_{iso}$(Å$^2$) |
|---|---|---|---|---|---|---|
| N1 | 16i | 0.803(11) | 0.063(8) | 0.0718(9) | 1.0 | 0.0678(8) |
| C1 | 16i | 0.617(11) | 0.076(8) | 0.015(1) | 1.0 | 0.0678(8) |
| C2 | 16i | 0.688(11) | 0.419(9) | -0.020(1) | 1.0 | 0.0678(8) |
| N2 | 16i | 0.514(12) | 0.466(8) | -0.078(1) | 1.0 | 0.0678(8) |
| H1 | 16i | 0.625(12) | 0.093(13) | 0.1072(9) | 1.0 | 0.0678(8) |
| H2 | 16i | 0.960(14) | 0.279(7) | 0.076(1) | 1.0 | 0.0678(8) |
| H3 | 16i | 0.340(13) | 0.025(7) | 0.020(2) | 1.0 | 0.0678(8) |
| H4 | 16i | 0.591(11) | 0.593(15) | 0.014(1) | 1.0 | 0.0678(8) |
| H5 | 16i | 0.724(11) | -0.094(11) | -0.020(1) | 1.0 | 0.0678(8) |
| H6 | 16i | 0.971(13) | 0.451(9) | -0.023(1) | 1.0 | 0.0678(8) |
| H7 | 16i | 0.263(14) | 0.551(9) | -0.076(2) | 1.0 | 0.0678(8) |
| H8 | 16i | 0.533(11) | 0.262(9) | 0.109(1) | 1.0 | 0.0678(8) |

| Se | 4e | 0.50000 | 0.50000 | 0.3142(1) | 1.0 | 0.063(1) |
|---|---|---|---|---|---|---|
| Fe | 4d | 0.00000 | 0.50000 | 0.25000 | 1.0 | 0.096(2) |

Space group: $I4/m$ (No. 86); $a$ = 3.8565(2) Å, $c$ = 21.4257(6) Å, $V$ = 318.66(3) Å$^3$, $U_{iso}$ is the isotropic displacement parameter.

**Table S3. Structural Parameters.** Atomic coordinates for ***C*** - $(C_2N_2H_8)Fe_2Se_2$ from NPD refinement at 3K. $R_p$ = 5.46% and $R_{wp}$ = 7.64%.

| Atom | Wyckoff site | x | y | z | Occupation | $U_{iso}$(Å$^2$) |
|---|---|---|---|---|---|---|
| N1 | 16h | -0.165(4) | 0.4588(8) | -0.679(4) | 1.0 | 0.021(2) |
| C1 | 16h | -0.036(3) | 0.4807(8) | -0.351(3) | 1.0 | 0.021(2) |
| C2 | 16h | -0.032(6) | 0.5508(8) | -0.302(2) | 1.0 | 0.021(2) |
| N2 | 16h | 0.106(7) | 0.5667(11) | 0.034(3) | 1.0 | 0.021(2) |
| H1 | 16h | -0.166(9) | 0.4110(8) | -0.688(6) | 1.0 | 0.021(2) |
| H2 | 16h | -0.011(7) | 0.4737(16) | -0.879(3) | 1.0 | 0.021(2) |
| H3 | 16h | 0.227(4) | 0.4630(13) | -0.311(3) | 1.0 | 0.021(2) |
| H4 | 16h | -0.192(4) | 0.4594(7) | -0.141(4) | 1.0 | 0.021(2) |
| H5 | 16h | -0.297(7) | 0.5686(8) | -0.354(2) | 1.0 | 0.021(2) |
| H6 | 16h | 0.134(7) | 0.5716(11) | -0.504(3) | 1.0 | 0.021(2) |
| H7 | 16h | -0.088(7) | 0.5783(22) | 0.203(5) | 1.0 | 0.021(2) |
| H8 | 16h | 0.269(10) | 0.6041(16) | 0.022(4) | 1.0 | 0.021(2) |
| Se | 4c | 0.50000 | 0.3166(3) | 0.75000 | 1.0 | 0.007(1) |
| Fe | 4b | 0.50000 | 0.25000 | 0.25000 | 1.0 | 0.011(3) |

Space group: *Cmcm* (No. 63); $a$ = 3.8779 (2) Å, $b$ = 21.359 (2) Å and $c$ = 3.8454 (4) Å, $V$ = 318.50(2) Å$^3$, $U_{iso}$ is the isotropic displacement parameter.